\documentclass[prb,twocolumn,superscriptaddress]{revtex4-1}

\usepackage[pdftex]{graphicx}
\usepackage{amsmath}
\usepackage{amsfonts}
\usepackage{amssymb}
\usepackage{bm}
\usepackage{amsbsy}
\usepackage{slashed}
\usepackage{mathtools}

\usepackage{pgfplots}
\usepackage{braket}
\usepackage{color}
\usepackage{hyperref}

\begin{document}

\title{Unified Fock space representation of fractional quantum Hall states}

\author{Andrea Di Gioacchino}
\affiliation{Dipartimento di Fisica, Universit\`a degli Studi di Milano, via Celoria 16, 20133 Milano, Italy}
\affiliation{INFN Milano, via Celoria 16, 20133 Milano, Italy}
\author{Luca Guido Molinari}
\affiliation{Dipartimento di Fisica, Universit\`a degli Studi di Milano, via Celoria 16, 20133 Milano, Italy}
\affiliation{INFN Milano, via Celoria 16, 20133 Milano, Italy}
\author{Vittorio Erba}
\affiliation{Dipartimento di Fisica, Universit\`a degli Studi di Milano, via Celoria 16, 20133 Milano, Italy}
\author{Pietro Rotondo}
\affiliation{School of Physics and Astronomy, University of Nottingham, Nottingham, NG7 2RD, UK}
\affiliation{Centre for the Mathematics and Theoretical Physics of Quantum Non-equilibrium Systems, University of Nottingham, Nottingham NG7 2RD, UK}

\begin{abstract}
Many bosonic (fermionic) fractional quantum Hall states, such as Laughlin, Moore-Read and Read-Rezayi wavefunctions, belong to a special class of orthogonal polynomials: the Jack polynomials (times a Vandermonde determinant). This fundamental observation allows to point out two different recurrence relations for the coefficients of the permanent (Slater) decomposition of the bosonic (fermionic) states. Here we provide an explicit Fock space representation for these wavefunctions by introducing a two-body squeezing operator which represents them as a Jastrow operator applied to reference states, which are in general simple periodic one dimensional patterns. Remarkably, this operator representation is the same for bosons and fermions, and the different nature of the two recurrence relations is an outcome of particle statistics.    
\end{abstract}

\maketitle

\section{Introduction}
Model wavefunctions, such as Laughlin \cite{Laughlin:PRL:1983}, Moore-Read \cite{Moore:NPB:1991} and Read-Rezayi \cite{Read:PRB:1999} states, together with the composite fermions picture \cite{Jain:PRL:1989,Jain:Book:2007}, describe with incredible accuracy the ground state at different filling fractions of strongly correlated two dimensional electrons in the fractional quantum Hall effect (FQHE) \cite{Tsui:PRL:1982}. 
They have an elegant representation as functions of the coordinates of the $N$ electrons. For instance, in the symmetric gauge 
and neglecting the Gaussian factor, the Laughlin state is a homogeneous polynomial of the variables $z_i = x_i - i y_i$: 
\begin{equation}
\psi_L^{(q)} (\vec{z}) = \prod_{j<i} (z_i-z_j)^q\,,
\end{equation}
where $\vec{z} = (z_1,z_2,\dots,z_N)$ is the  vector of particle positions and the positive integer $q$ is related to the filling fraction $\nu$ 
by   $\nu = 1/q$.  The state is bosonic (fermionic) if $q$ is even (odd). 

Despite their simplicity as functions of the coordinates, Laughlin wavefunctions are non-trivial superpositions of permanents (Slater determinants) of the
single-particle states of the lowest Landau level (LLL), whose number increases exponentially with $N$.  
The problem of finding the expansion of a Laughlin state on this single-particle basis was considered a formidable task for a long time \cite{Yoshioka:Book:2013}. In the nineties it was addressed by Dunne \cite{Dunne:IJMP:1993} and by Di Francesco and coworkers \cite{Difrancesco:IJMPA:1994}. They found that the coefficients of the Slater decomposition of the $q=3$ Laughlin state possess symmetries, suggesting that a deep mathematical structure is hidden. 

Recently Haldane and Bernevig \cite{Bernevig:PRL:2008} shed light on this structure, by noticing that many bosonic FQHE states belong to a special class of symmetric orthogonal polynomials: the Jack polynomials $J_{\bm{\lambda}}^\alpha (\vec z)$, widely studied in the mathematical literature \cite{Macdonald:OUP:1995,Stanley:AM:1989,Feigin:IMRN:2002}. The authors were then able to find a recurrence relation satisfied by the coefficients of the permanent decomposition of the bosonic states. A different recurrence 
relation was recognized to hold for fermionic states, which are the product of a Jack polynomial and the Vandermonde determinant 
\cite{Bernevig:PRL:2009,Thomale:PRB:2011}. 

On another side, Bergholtz and coworkers \cite{Nakamura:PRL:2012} considered a \emph{truncation} of the Trugman-Kivelson hamiltonian, which is known to have the Laughlin state  $q=3$ as ground state at $\nu=1/3$ \cite{Trugman:PRB:1985}. They showed that the ground state of this approximated problem has the Fock space representation
\begin{equation}
|\psi_{GS}\rangle = U |{100100100\cdots}\rangle\,,
\label{bergholtz}
\end{equation}
where $U$ is a two-body fermionic operator and the reference state is the corresponding $\nu=1/3$ thin-torus occupancy pattern \cite{Rotondo:PRL:2016, Bergholtz:PRL:2005, Bergholtz:PRL:2007}. Remarkably, the precise form of $U$ is known. This is not the case for other FQHE states; however the link between fractional quantum Hall states and Jack polynomials suggests that a similar explicit representation should exist.  

This is precisely the result of this paper: we exhibit an explicit Fock space representation of the bosonic (fermionic) fractional quantum Hall states that can be written as a Jack polynomial (times a Vandermonde determinant). More in detail, we show that these wavefunctions result from the action of a \emph{universal} operator acting on proper ``root states'', in analogy with Eq.~\eqref{bergholtz}. Remarkably, we find that this operator has the same functional form for both the bosonic and fermionic sectors, thus bringing together the two recurrence relations found previously\cite{Thomale:PRB:2011}, whose difference is shown to descend from particle statistics.   

The paper is organized as follows. In Sec. \ref{sec_not}, we introduce a coherent formalism to properly treat the many-particle problem in the LLL, both in the abstract Fock space and in its coordinate representation space (Bargmann space). We introduce the standard creation and destruction operator algebra, and use it to implement the \emph{squeezing operation}, fundamental in the theory of Jack polynomials. In Sec. \ref{sec_main} we discuss our main result, i.e. the Fock space representation of many fractional quantum Hall model states, and we recover in a novel and unified way the known recurrence relations between the coefficients of the state decompositions on the bases introduced in Sec. \ref{sec_not}.
Finally, in Sec. \ref{sec_conc} we give our conclusions.

\section{Squeezings in Fock space}\label{sec_not}

In the symmetric gauge the single particle basis of the LLL are the functions $u_\lambda (z) = z^\lambda / (\lambda !)$ where $z = x - i y$ is the particle's position (in unit of magnetic length) and $\lambda = 0,1,2,\dots $ is the angular momentum 
(the Gaussian factor is included in the Hilbert space measure).
They build the basis of permanents (determinants) for $N$ bosons (fermions).

\subsection{Basis for the Fock space of N particles}

Let $\ket{\lambda}$, $\lambda= 0, 1, 2, \dots $ be an orthogonal basis for a single particle Hilbert space, where each vector may be not normalized. 
The associated normalized basis is  $\tfrac{1}{\sqrt{\nu_\lambda}} \ket{\lambda}$, with $\nu_\lambda = \braket{\lambda | \lambda}$.\\
An orthogonal basis for the Hilbert space of $N$ bosons or fermions is given respectively by:
\begin{equation}
\begin{split}
& \ket{\textrm{per}_{\bm{\lambda}}} = \sum_{\pi} \ket{\lambda_{\pi_1} \dots  \lambda_{\pi_N}}, \\
& \ket{\textrm{sl}_{\bm{\lambda}}} = \sum_{\pi} (-)^{\pi} \ket{\lambda_{\pi_1} \dots  \lambda_{\pi_N}},
\end{split}
\end{equation}
where ${\bm{\lambda}} = (\lambda_1, \dots, \lambda_N)$, with $\lambda_1 \geq \dots \geq \lambda_N$, specifies the single particle states (for fermions equality is forbidden), the sum runs over the permutations of indices $\{1, \dots, N\}$ and $(-)^{\pi}$ is the parity of the permutation. 
In accordance with the mathematical literature \cite{Macdonald:OUP:1995,Stanley:AM:1989}, the sequence ${\bm{\lambda}} $ is called ``a partition'' (of length $N$).\\
For example: 
\begin{equation}\label{ex}
\begin{split}
& \ket{\textrm{per}_{440}} = 2\ket{440} + 2\ket{404} + 2\ket{044}, \\
& \ket{\textrm{sl}_{410}} = \ket{410} - \ket{401} + \ket{041} - \ket{014} + \ket{104} - \ket{140}.
\end{split}
\end{equation}
These bases can be normalized by a factor which accounts for the normalization of single particle states and for the multiplicities produced by permutations:
\begin{equation}
\frac{1}{\sqrt{\nu_{\bm{\lambda}}}} = \sqrt{\frac{1}{N!}} \sqrt{\frac{1}{n_0! \dots n_{\infty}!}} \sqrt{\frac{1}{\nu_{\lambda_1} \dots \nu_{\lambda_N}}},
\end{equation}
where $n_\lambda$ is the number of repetitions of $\lambda$ in the partition ${\bm{\lambda}}$. Notice that for fermions, $n_i = 0, 1$.\\
The bases have two equivalent notations (see Fig. 1a):
\begin{itemize}
	\item the single particle notation,  where the emphasis is on the single-particle quantum numbers $\lambda_i$, \mbox{$\ket{\lambda_1 \geq \lambda_2 \geq \dots \geq \lambda_N}$}; 
	\item the occupation number notation, where the emphasis is on how many particles share the same quantum number, $\ket{n_0, n_1, \dots, n_{\infty}}$.
\end{itemize}

\subsection{Creation and destruction operators}

We introduce a canonical algebra of creation and destruction operators for the 1-particle orthonormal states $\frac{1}{\sqrt{\nu_\lambda}} \ket{\lambda}$, for bosons and fermions:

\begin{equation}
\begin{split}
a_\lambda^\dagger \ket{\dots n_\lambda \dots} &= (\pm 1)^{n_0 + \dots + n_{\lambda-1}} \sqrt{n_\lambda + 1} \ket{\dots n_\lambda+1 \dots}, \\
a_\lambda \ket{\dots n_\lambda \dots} &= (\pm 1)^{n_0 + \dots + n_{\lambda-1}} \sqrt{n_\lambda} \ket{\dots n_\lambda-1 \dots},\\ 
a_\lambda \ket{\dots n_\lambda \dots} &= 0 \quad \text{if } n_\lambda = 0. \\
\end{split}
\end{equation}
where + is for bosons and - for fermions. \\
From now on, we will use Latin letters to indicate single-particle quantum numbers, Greek letters to indicate partitions and Greek letters with subscripts to indicate the elements of the corresponding partition.
The explicit actions on $\ket{\textrm{per}_{\bm{\lambda}}}$ are
\begin{equation}
\begin{split}
a_r^\dagger \ket{\textrm{per}_{\bm{\lambda}}} &= \sqrt{\tfrac{1}{(N+1) \, \nu_r}} \, \ket{\textrm{per}_{{\bm{\lambda}} +r}}, \\
a_r \ket{\textrm{per}_{\bm{\lambda}}} &= n_r \sqrt{N \, \nu_r} \, \ket{\textrm{per}_{{\bm{\lambda}} -r}} \qquad \text{or 0 if } r \notin {\bm{\lambda}},
\end{split}
\end{equation}
where ${\bm{\lambda}} \pm r$ is the partition obtained by adding or removing the entry $r$ in ${\bm{\lambda}}$.
The action on $\ket{\textrm{sl}_{\bm{\lambda}}}$ is the same, except for the $(-)^{n_0 + \dots + n_{r-1}}$ factor (remember that $n_i = 0, 1$ for fermions).\\
The number operator is $\hat{n}_\lambda = a^\dagger_\lambda a_\lambda$.

\subsection{Squeezing operations}\label{sec_squeez}
The creation and destruction operators allow for an efficient description of the \emph{squeezing operation}, ubiquitous in the theory of Jack polynomials.
Squeezing operations are implemented by the following operator:
\begin{equation}\label{sqop}
s_{u, m, k} = \sqrt{\frac{\nu_{u-k} \nu_{m+k}}{\nu_u \nu_m}} a^\dagger_{u+k} a^\dagger_{m-k} a_m a_u 
\end{equation}
for $0 \leq u < m$, $0 < k < m-u$.
Notice that this operator satisfies $s_{u, m, k} = \pm s_{u, m, m-u-k}$, based on the statistics of the particles. \\
The explicit action of operator \eqref{sqop} on the $\ket{\textrm{per}_{\bm{\lambda}}}$ basis is:
\begin{equation}\label{sonper}
\begin{split}
s_{u, m, k} \ket{\textrm{per}_{\bm{\lambda}} }
=
n_{u} n_{m} \ket{\textrm{per}_{\bm{\mu}}} 
\end{split}
\end{equation}
where ${\bm{\mu}}$ is constructed from ${\bm{\lambda}}$ by substituting two particles of  quantum numbers $u$ and $m$ with two particles of  quantum numbers $u+k$ and $m-k$, hence by ``squeezing the two particles by $k$''. \\
On $\ket{\textrm{sl}_{\bm{\lambda}}}$, one has
\begin{equation}
\begin{split}
s_{u, m, k} \ket{\textrm{sl}_{\bm{\lambda}}} 
= (-)^{N_\mathrm{sw}}\ket{\textrm{sl}_{\bm{\mu}}},
\end{split}
\end{equation}
where $N_\mathrm{sw}$ is the number of exchanges that restore the decreasing order of the sequence.

Squeezing operations can be used to introduce a partial ordering on partitions: ${\bm{\lambda}} > {\bm{\mu}} \Leftrightarrow$ $\ket{\textrm{per}_{{\bm{\mu}}}}$ can be constructed from $\ket{\textrm{per}_{\bm{\lambda}}}$ through squeezing operations (and same with $\textrm{sl}$ in the fermionic case). 
The notation ${\bm{\mu}} \leftarrow {\bm{\lambda}}$ is used if $\ket{\textrm{per}_{{\bm{\mu}}}}$ or $\ket{\textrm{sl}_{{\bm{\mu}}}}$ is obtained with a single squeezing operation from $\ket{\textrm{per}_{\bm{\lambda}}}$ or $\ket{\textrm{sl}_{\bm{\lambda}}}$.

\begin{figure}[t]
	\includegraphics[width=\linewidth,keepaspectratio]{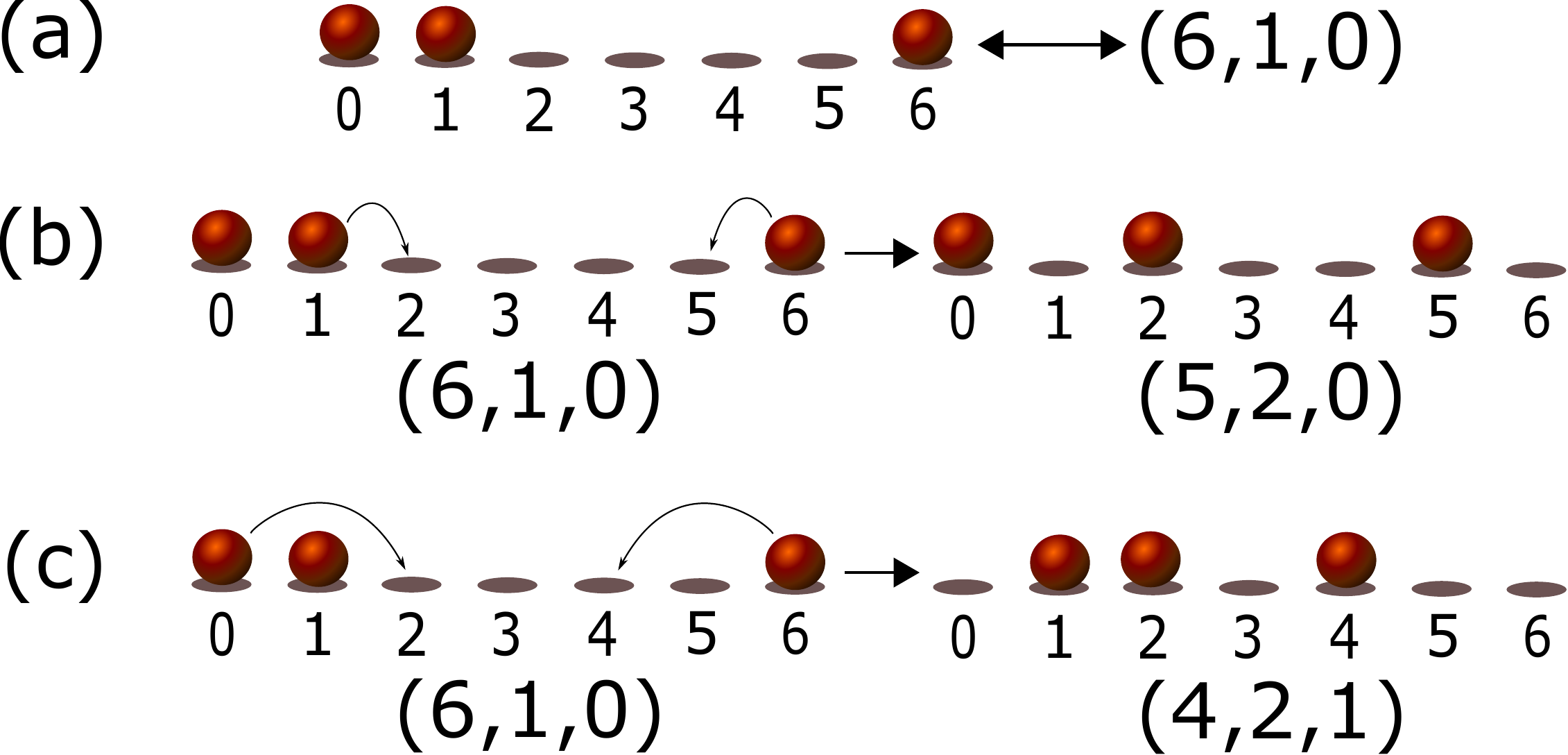}
	\caption{(a) Each site $0,1,2,\dots $ is an angular momentum state in the LLL. A configuration of occupancies corresponds to a partition. Here the lattice configuration on the left corresponds to the partition $(6,1,0)$. (b) Lattice and partition representation of one of the possible squeezings of the partition $(6,1,0)$. In Fock space, this corresponds to the action of the two-body operator $a^{\dagger}_2 a^{\dagger}_5 a_6 a_1$. (c) The squeezing $(6,1,0)\rightarrow (4,2,1)$ shows the power of the Fock space representation. Indeed using the operator $a^{\dagger}_2 a^{\dagger}_4 a_6 a_0$, minus signs arising in the fermionic case are automatically taken into account, making the unification of the bosonic and fermionic FQH states made in Eq. \eqref{res} possible.} \label{figsqueeze}
\end{figure}

Notice that a squeezing does not change the quantity $\sum_{i=1}^N  \lambda_i$. In the quantum Hall effect  the squeezing operations preserve the total angular momentum of the system.
Two examples are given in Fig. \ref{figsqueeze}.

\subsection{Bargmann space representation}\label{sec_barg}

Within this formalism, we recover the usual LLL many-particle wavefunctions in the Bargmann space, in which we consider the basis of monomials $\braket{z|r} = z^r$, $r= 0, 1, 2, \dots $, with $\nu_r = r!$:
\begin{equation}
\begin{split}
&\braket{\vec{z}|\textrm{per}_{\bm{\lambda}}}   = \textrm{per}_{\bm{\lambda}}(\vec{z}) = \text{Permanent}[z_i^{\lambda_j}], \\
&\braket{\vec{z}|\textrm{sl}_{\bm{\lambda}}}   = \textrm{sl}_{\bm{\lambda}}(\vec{z}) = \text{Determinant}[z_i^{\lambda_j}].
\end{split}
\end{equation}
For example, the states in Eq. \eqref{ex} become:
\begin{equation}
\begin{split}
&\textrm{per}_{440}(z_1, z_2, z_3) = 2 z_1^4z_2^4 + 2 z_1^4z_3^4 + 2 z_2^4z_3^4, \\
&\textrm{sl}_{410}(z_1, z_2, z_3) \! = \! z_1^4z_2 \! - \! z_1^4z_3 \! + \! z_2^4z_3 \! - \!  z_2z_3^4 \! + \! z_1z_3^4 \! - \! z_1z_2^4.\\
\end{split}
\end{equation}
Notice that while $\textrm{sl}_{\bm{\lambda}}$ coincides with the antisymmetric monomials used in mathematics, $\textrm{per}_{\bm{\lambda}}$ does not coincide with the usual symmetric monomials $\textrm{m}_{\bm{\lambda}}$. The latter are defined as $\textrm{per}_{\bm{\lambda}}$, with only one copy of each monomial.
For example:
\begin{equation}
\begin{split}
\textrm{m}_{440}(z_1, z_2, z_3) &= \tfrac{1}{2}\textrm{per}_{440}(z_1, z_2, z_3).
\end{split}
\end{equation}
In general, $\textrm{m}_{\bm{\lambda}} (\vec{z}) = \frac{1}{n_0! \dots n_\infty!} \textrm{per}_{\bm{\lambda}} (\vec{z})$. Thus, the normalization constant for $\textrm{m}_{\bm{\lambda}}$ is
\begin{equation}
\frac{1}{\sqrt{\nu_{\bm{\lambda}}^\textrm{m} }} = \sqrt{\frac{1}{N!}} \sqrt{\frac{n_0! \dots n_{\infty}!}{\nu_{\lambda_1} \dots \nu_{\lambda_N}}}.
\end{equation}

It is useful to specify the action of $a^\dagger$, $a$, and $s$ on the basis of monomials:
\begin{equation}\label{sonmon}
\begin{split}
& a_r^\dagger \textrm{m}_{\bm{\lambda}} = \sqrt{\tfrac{1}{(N+1) \, \nu_r}} \, (n_r+1) \, \textrm{m}_{{\bm{\lambda}} + r} ,\\
& a_r \textrm{m}_{\bm{\lambda}} = \sqrt{N \, \nu_r} \,  \textrm{m}_{{\bm{\lambda}} -r} \qquad \text{or 0 if } r \notin {\bm{\lambda}} ,\\
& \hat{n}_r \textrm{m}_{\bm{\lambda}} = n_r \textrm{m}_{\bm{\lambda}}, \\
& s_{u, m, k} \textrm{m}_{\bm{\lambda}} = 
	\begin{cases}
	0 & \!\!\!\!\!\!\!\!\! \text{if $u\notin {\bm{\lambda}}$ or $m\notin {\bm{\lambda}}$}\\
	(n_{u+k}+2)(n_{u+k}+1) \textrm{m}_{{\bm{\mu}}}& \; k = \frac{m-u}{2} \in \mathbb{N}\\
	(n_{u+k}+1)(n_{m-k}+1) \textrm{m}_{{\bm{\mu}}}& \; \text{otherwise},
	\end{cases}
\end{split}
\end{equation}
where ${\bm{\mu}}$ is the partition squeezed from ${\bm{\lambda}}$.

\section{Fock space representation of FQHE Jack states }\label{sec_main}
\subsection{Main result}
The bosonic (fermionic) Jack polynomials (times a Vandermonde determinant), including Laughlin, Moore-Read and Read-Rezayi wavefunctions, can be obtained in Fock space by the action of an operator on an appropriate root: 
\begin{equation}\label{res}
|\psi_{\bm{\lambda}} \rangle = \left[ \mathbb{I} - \frac{K}{E_{\bm{\lambda}} - D} S \right ]^{-1}  |{\bm{\lambda}} \rangle .
\end{equation}
The root $\ket{{\bm{\lambda}}} $ is a permanent $\ket{\textrm{per}_{\bm{\lambda}}}$ (bosons) or a determinant $\ket{\textrm{sl}_{\bm{\lambda}}}$ (fermions). In the following, we will unify the notation and use simply $\ket{{\bm{\lambda}}} $.
$S$  is a two-body operator,
which we call \emph{squeezing operator}:
\begin{equation}\label{squeezing_op}
S=\sum_{s=0}^{\infty} \sum_{t=1}^{\infty} \sum_{u=1}^{\infty} u \sqrt{\frac{(s+t)! (s+u)!}{(s+t+u)! s!}} \: a^\dagger_{s+t} \: a^\dagger_{s+u} \: a_{s+t+u} \: a_{s}\,.
\end{equation}

\begin{figure}[t]
	\includegraphics[width=\linewidth,keepaspectratio]{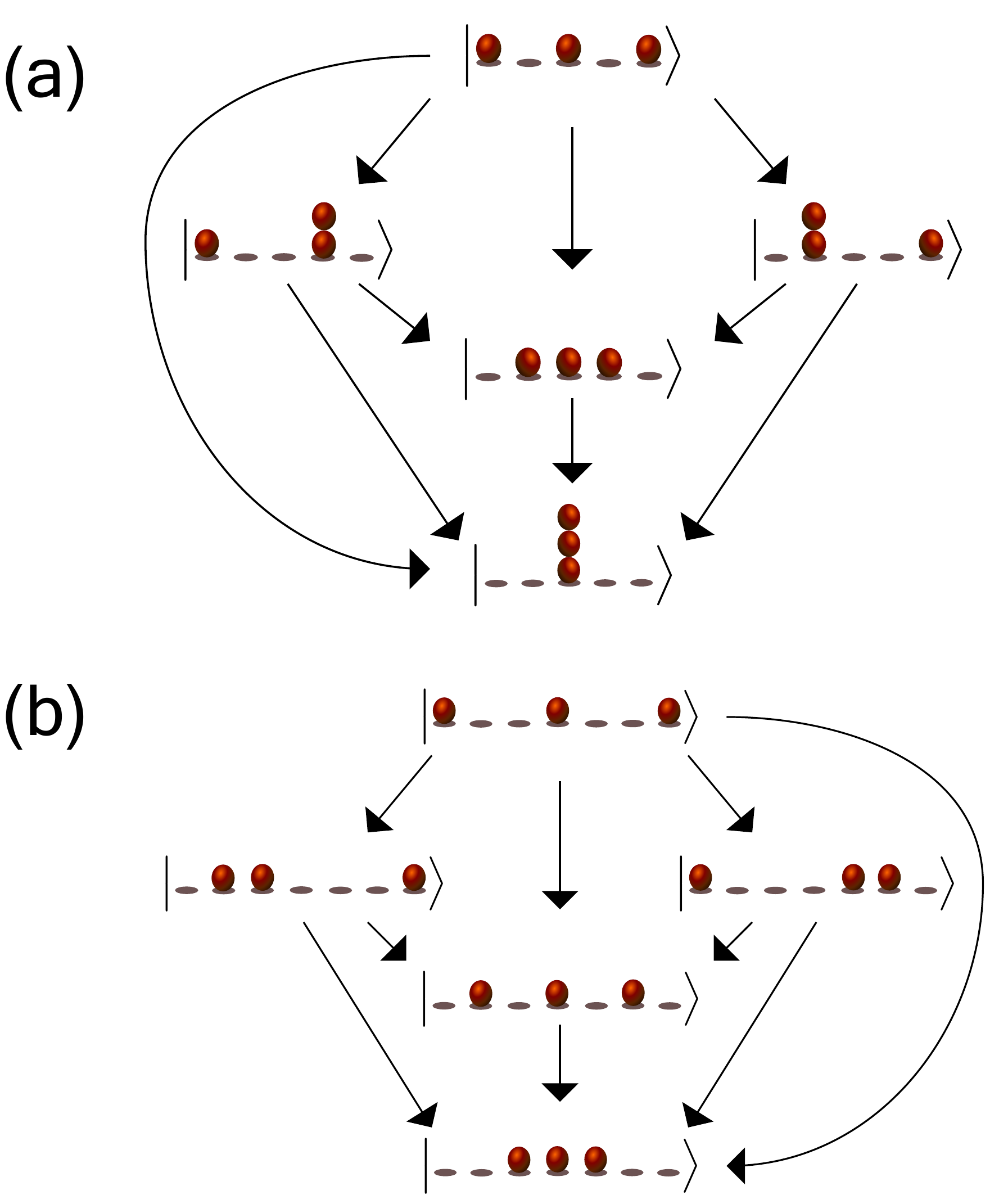}\label{fig:bose_chain}
	\caption{The repeated action of the squeezing operator of Eq. \eqref{squeezing_op} on the bosonic and fermionic Laughlin states with $\nu=1/2$ (a) and $\nu=1/3$ (b), with three particles. Each state is specified by the occupancy numbers, and each arrow represents the action of a squeeze. The operator in Eq. \eqref{squeezing_op} automatically takes into account the possibility of multiple occupancies of the bosonic case.}\label{fig:squeeze_chain}
\end{figure}

The sums encompass all possible squeezing operations (``squeezings'' for brevity) described in section \ref{sec_squeez}, where one particle in $s$ and one in $s+t+u$ are taken to $s+t$ and $s+u$.
The initial distance $t+u$ is squeezed to $|t-u|$. The coefficients produce the proper combination of squeezed permanents (determinants) that compound a Jack polynomial (times a Vandermonde).  A simple example
is illustrated in Fig. \ref{fig:squeeze_chain}. \\
$D$ is an operator diagonal on permanents or determinants: $D |{\bm{\mu}}\rangle  = E_{\bm{\mu}} |{\bm{\mu}}\rangle $ with 
\begin{align}\label{eq_eigenval}
E_{\bm{\mu}} = \bra{{\bm{\mu}}} D \ket{{\bm{\mu}}} = \sum_{j=1}^N \mu_j^2 + \frac{K}{2} \sum_{i<j} (\mu_i-\mu_j),
\end{align}
where ${\bm{\mu}} =(\mu_1,\dots,\mu_N)$ 
and $K = (1-2q\pm 1)/(p+1)$ ($+1$ for bosons, $-1$ for fermions). The positive integers $p,q$ are related to the filling fraction of the wavefunction through $\nu = p/q$ \cite{Thomale:PRB:2011}. In the literature, the parameter $\alpha $ is often used: $K=2/\alpha$ for bosons and $K=(2/\alpha)-2$ for fermions. $\alpha$ is the custom label of Jack polynomials, $J_{\bm{\lambda}}^\alpha (\vec z)$.

The well known series of bosonic and fermionic FQH states are recovered if the roots are chosen as permanents (B) or determinants (F), with partitions specified by the occupation numbers
\begin{align}
& {\bm{\lambda}}^{(B)} = (p,\underbrace{0,\dots,0}_{q-1},p,\underbrace{0,\dots,0}_{q-1},p,\dots) \label{root}, \\
& {\bm{\lambda}}^{(F)} = (\underbrace{1,\dots,1}_{p},\underbrace{0,\dots,0}_{q-p},\underbrace{1,\dots,1}_{p},\underbrace{0,\dots,0}_{q-p},\dots) \label{root2}.
\end{align}
The choice $p=1$ gives the Laughlin states $q$ (bosonic (B) partition for even $q$, fermionic partition (F) for odd $q$), the partitions with $p=2$, $q=2m+2$ give the $\nu=2/(2m+2)$ Moore-Read states (bosonic (fermionic) for even (odd) $m$) and the Read-Rezayi $k$ states are obtained with the bosonic partition with $p=k$, $q=2$ \cite{Baratta:NPB:2011}.

\subsection{Proof of main result}
The bosonic (B) and fermionic (F) wavefunctions $\psi_{\bm{\lambda}} (z_1,\dots ,z_N) $ are eigenstates of the ``Generalized Laplace-Beltrami'' operator  \cite{Thomale:PRB:2011}, which acts on the Bargmann space introduced in section \ref{sec_barg}:
\begin{align}
H=\sum_{k=1}^N (z_k\partial_k)^2  + K \sum_{j\neq k} \frac{z_k+ z_j}{z_k-z_j} \left (z_k\partial_k - z_j\partial_j\right)\\
- K\sum_{j\neq k} \frac{z_j^2+z_k^2}{(z_j-z_k)^2}(1-\pi_{jk}) \nonumber
\end{align}
where $\pi_{jk}$ is the exchange operator of particles $j,k$. In the bosonic (fermionic) sector it is
$\pi_{jk}=1$ ($-1$). By writing $H$ in second quantization we obtained the form (the detailed calculations are in appendix \ref{AppA})
\begin{equation}
H=D + K S, 
\end{equation}
where $S$ is the operator in Eq. \eqref{squeezing_op} and $D$ is the same operator of Eq. \eqref{eq_eigenval}. It incorporates the kinetic term and the diagonal part of the two-body potential, and its explicit form is (up to an irrelevant additive constant for fermions):
\begin{align}
D = \sum_{m=0}^\infty m^2 a^\dagger_m a_m + \frac{K}{2} \sum_{m'<m} (m-m') a^\dagger_m a_m a^\dagger_{m'} a_{m'}.
\end{align}
A notable feature of the formalism of second quantization is to be independent of the number $N$ of particles.\\
For any state $|{\bm{\lambda}} \rangle $ of length $N$ there is a power $n$ such that $S^n|{\bm{\lambda}} \rangle =0$. 
This allows to prove the following proposition: {\em If $E_{\bm{\lambda}} $ is a non degenerate eigenvalue of $D$, with eigenvector $|{\bm{\lambda}}\rangle $, then
\begin{align}
|\psi_{\bm{\lambda}} \rangle =  \sum_{k=0}^n [(E_{\bm{\lambda}} -D)^{-1} S]^k |{\bm{\lambda}}\rangle
\label{prop} 
\end{align}
is eigenvector of $H$ with the same eigenvalue $E_{\bm{\lambda}} $.}\\
Indeed multiplication of Eq. \eqref{prop} by $(E_{\bm{\lambda}} -D)^{-1}S$  gives 
\begin{equation}
(E_{\bm{\lambda}} -D)^{-1}S |\psi_{\bm{\lambda}} \rangle = |\psi_{\bm{\lambda}} \rangle - |{\bm{\lambda}} \rangle\,.  \label{eq_for_psi}
\end{equation}
Multiplication by $(E_{\bm{\lambda}} -D)$ gives $S|\psi_{\bm{\lambda}} \rangle = (E_{\bm{\lambda}} -D)|\psi_{\bm{\lambda}} \rangle$.\\
The reverse is also true: if $|\psi\rangle $ is an eigenstate of $H$ with non-degenerate eigenvalue $E$, then 
$|{\bm{\lambda}} \rangle = |\psi\rangle -(E-D)^{-1} S |\psi \rangle $ is an eigenvector of $D$ with same eigenvalue.\\

\subsection{Recovering the recurrence relations}

By construction \eqref{res} implies that $|\psi_{\bm{\lambda}} \rangle $ is a linear combination of permanents (determinants) that are obtained by squeezings 
on the root $|{\bm{\lambda}} \rangle $:
\begin{align}
|\psi_{\bm{\lambda}} \rangle = \sum_{{\bm{\mu}}\le {\bm{\lambda}}} b_{{\bm{\lambda}} ,{\bm{\mu}}} |{\bm{\mu}}\rangle   \label{eq_rec}
\end{align}
with coefficients $b_{{\bm{\lambda}},{\bm{\mu}}} = \langle {\bm{\mu}}|\psi_{\bm{\lambda}} \rangle $.
Eq. \eqref{eq_for_psi} gives $b_{{\bm{\lambda}}, {\bm{\lambda}}}=1$ and, straightforwardly,  the recursive relation for ${\bm{\mu}}\neq{\bm{\lambda}} $: 
\begin{align} 
b_{{\bm{\lambda}},{\bm{\mu}}} = \frac{K}{E_{\bm{\lambda}} -E_{\bm{\mu}}}\sum_{{\bm{\mu}}\leftarrow{\bm{\theta}}<{\bm{\lambda}} } \langle {\bm{\mu}}|S|{\bm{\theta}}\rangle  b_{{\bm{\lambda}},{\bm{\theta}}} \label{eq_recursion}
\end{align}
The notation ${\bm{\mu}}\leftarrow{\bm{\theta}}<{\bm{\lambda}}$ means that the sum only involves partitions ${\bm{\theta}}$ that yield ${\bm{\mu}}$ after just one squeezing (${\bm{\mu}}\leftarrow{\bm{\theta}}$) and that descend from the root ${\bm{\lambda}} $ after one or more squeezings (${\bm{\theta}}<{\bm{\lambda}}$).
All partitions involved have the same length $N$ and angular momentum of the root, $\sum_{j=1}^N \lambda_j$. In the difference of eigenvalues (written in Eq.\eqref{eq_eigenval}) in Eq.\eqref{eq_recursion} the constant term for fermions cancels. For both statistics, one evaluates :
\begin{equation}
E_{\bm{\lambda}}-E_{\bm{\mu}} = \sum_{j=1}^N  (\lambda_j-\mu_j)(\lambda_j+\mu_j -  jK )
\end{equation}
In appendix \ref{AppB} we show that the sums on partitions in Eq.\eqref{eq_recursion} and in $S$ give:
\begin{align}\label{stp2}
&\sum_{{\bm{\mu}}\leftarrow{\bm{\theta}} <{\bm{\lambda}} } \langle {\bm{\mu}}| S |{\bm{\theta}} \rangle  b_{{\bm{\lambda}},{\bm{\theta}}} \\
&= \sum_{{\bm{\mu}} \leftarrow {\bm{\theta}} <{\bm{\lambda}}}  
b_{{\bm{\lambda}},{\bm{\theta}}} \times \begin{cases} (\theta_i-\theta_j)  y_{\theta_i,\theta_j} & \text{(bosons)} \\  (\mu_i-\mu_j) (-1)^{N_\mathrm{sw}} &\text{(fermions)}  \end{cases}
\nonumber
\end{align}
where the partition $\bm{\mu}$ is obtained by a squeezing from $\bm{\theta}$, and this squeeze moves the particles with quantum numbers $\mu_i$ and $\mu_j$ in $\theta_i$ and $\theta_j$; $N_\mathrm{sw}$ is the number of swaps to properly reorder the partition after the squeeze and 
\begin{equation}\label{ydefper}
y_{\theta_i,\theta_j} = n_{\theta_i} n_{\theta_j}
\end{equation}
($n_{\theta_i}$ is the number of occurrences of $\theta_i$ in the partition ${\bm{\theta}} $).
The square root factors in Eq. \eqref{squeezing_op} are cancelled by the action of the creation and annihilation operators on the many-particle states. 

\begin{figure}[t]
	\includegraphics[width=\linewidth]{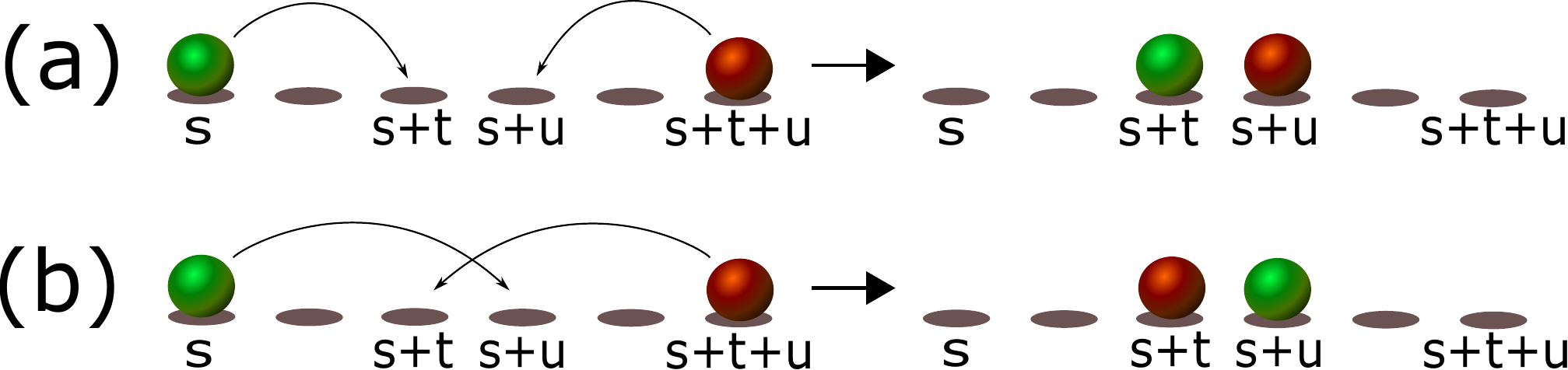}
	\caption{The different factors $\mu_i-\mu_j$ (bosons) and $\theta_i - \theta_j$ (fermions) in Eq. \eqref{stp2} can be understood graphically: squeezings (a) and (b) bring to the same physical state, because the two particles which are coloured in green and red are actually indistinguishable. However, due to the fact that in (b) there is a crossing between particles, the two states differ for a phase which depends on the statistics: $ + 1$ for bosons, $-1$ for fermions. The squeezing operator defined in Eq. \eqref{squeezing_op} automatically accounts for the difference between bosons and fermions, as it is explicitly shown in appendix \ref{AppB}.}
	\label{fig:doublesqueeze}
\end{figure}

The different factors $(\mu_i-\mu_j)$ or $(\theta_i -\theta_j)$ for the matrix element $\langle {\bm{\mu}}|S|{\bm{\theta}}\rangle $ are explained as a 
consequence of statistics. Consider two occupied sites $s$ and $s+t+u$ of the partition ${\bm{\theta}} $, and two occupied sites $s+t$ and $s+u$
of the partition ${\bm{\mu}}$, with $t\le u$ (see Fig. \ref{fig:doublesqueeze}).
There are two squeezings in $S$ connecting the pairs: the one making the particle in $s$ jump to $s+t$ and $s+t+u$ to $s+u$ (Fig. 3a), and the one that makes $s$ jump to $s+u$ and $s+t+u$ to $s+t$ (Fig. 3b). They differ by the exchange of $t$ and $u$, and give the same state because particles are indistinguishable. 
In the latter squeezing, the creation operators must be (anti)commuted. Therefore according to particle statistics the final coefficients for the pair of squeezings are $u \pm t$ where the upper (lower) sign holds for bosons (fermions). However: $u+t$ is the distance between particles before the squeezing, i.e. $\theta_i-\theta_j$, while $u-t$ is the distance after the squeezing, i.e. $\mu_i-\mu_j$.\\
This qualitative argument is behind the calculations done in appendix \ref{AppB}.

From Eqs. \eqref{eq_recursion} and \eqref{stp2} it is straightforward to obtain the recurrence relations for the bosonic and fermionic FQH states. We remark a technical point: in the mathematical literature on Jack polynomials in $N$ variables, the expansion is done on the monomial basis, introduced in section \ref{sec_barg} (we use sans serif symbols when referring to the monomial basis):
\begin{equation}\label{expansion}
\psi_{\bm{\lambda}} (\vec z) = m_{\bm{\lambda}} (\vec z)+ \sum_{{\bm{\mu}}<{\bm{\lambda}}} \mathsf{b_{{\bm{\lambda}}{\bm{\mu}}}} \, m_{\bm{\mu}} (\vec z),
\end{equation}
where $m_{\bm{\mu}} (\vec z)$ are monomials labelled by partitions ${\bm{\mu}}$ of same length $N$ as the root ${\bm{\lambda}} $. 
Since each monomial of a partition with at least one repeated element differs from the corresponding permanent by a constant factor, a change of basis results in a change of recurrence relations between coefficients. In particular, the difference is encoded in the $y$ factor (as discussed in appendix \ref{AppB}).
In the monomial basis, we obtain the following recurrence relations: 

\begin{align}
&\sf b^{(B)}_{{\bm{\mu}} {\bm{\lambda}}} = \frac{2/\alpha}{\rho^{(B)}_{{\bm{\lambda}}}-\rho^{(B)}_{\bm{\mu}}} \sum_{{\bm{\mu}}\leftarrow {\bm{\theta}} \leq {\bm{\lambda}} } b^{(B)}_{{\bm{\theta}} {\bm{\lambda}}} (\theta_i - \theta_j)  y_{\mu_i,\mu_j};\label{rec_rel_bose} \\
&\sf b^{(F)}_{{\bm{\mu}} {\bm{\lambda}}}=\frac{2(\frac{1}{\alpha}-1)}{\rho^{(F)}_{{\bm{\lambda}}} - \rho^{(F)}_{\bm{\mu}}} \sum_{{\bm{\mu}} \leftarrow{\bm{\theta}} \leq {\bm{\lambda}}} b^{(F)}_{{\bm{\theta}} {\bm{\lambda}}} (\mu_i - \mu_j) (-1)^{N_\mathrm{sw}},  \label{rec_rel_fermi}
\end{align}
where $\rho^{(B)}_{\bm{\lambda}}=\sum_i\lambda_i(\lambda_i - \frac{2}{\alpha} i)$, $\rho^{(F)}_{\bm{\lambda}} = \sum_i \lambda_i (\lambda_i + 2i (1-1/\alpha)) $ and
\begin{equation}\label{ydefmon}
\sf y_{\mu_i,\mu_j} = 
\left\{ 
\begin{aligned}
& \binom{n_{\mu_i}}{2} \: \, \, \quad \textrm{ if $\mu_i=\mu_j$ };\\
& n_{\mu_i} n_{\mu_j} \: \, \: \textrm{otherwise}.\\
\end{aligned} \right.
\end{equation}
These equations were firstly obtained by exploiting known properties of Jack polynomials \cite{Lapointe:JC:2001,Dumitriu:JSC:2007} and extending them to Jack polynomials times a Vandermonde \cite{Thomale:PRB:2011}.

\section{Conclusions}\label{sec_conc}
We show that the Fock space operators provide a natural formulation of the squeezing operations, ubiquitous in the mathematical and physics literature on Jack polynomials. We exhibit an explicit Fock space representation of bosonic (fermionic) FQH states of the form of a Jack polynomial (times a Vandermonde). Though the recurrence relations for bosons\cite{Bernevig:PRL:2008} and fermions\cite{Bernevig:PRL:2009} are different when expressed in the Bargmann space (Eqs.\eqref{rec_rel_bose} and \eqref{rec_rel_fermi}), our representation shows that this is only a result of particle statistics, thus allowing to treat bosons and fermions on the same footing. An open question remains on the reason why such Jack FQHE states are so effective for FQHE. Another interesting question to address, is to understand whether there is a connection between our Fock space representation and the matrix product state representation for the Laughlin wavefunction\cite{Iblisdir:PRL:2007}.

We finally stress that our results might be of interest in the context of integrable models, such as the generalized Calogero-Sutherland model \cite{Baker:NPB:1997, Kuramoto:book:2009}. To our knowledge, the Fock space approach introduced here, has not yet been considered in the literature on Jack polynomials.

\appendix

\section{Second quantization of generalized Laplace-Beltrami operator}\label{AppA}
Here, we use the Fock space formalism to rewrite the Laplace-Beltrami operator and his fermionic generalization in terms of creation and annihilation operators.

\subsection{Laplace-Beltrami operator and Jack polynomials}

In the Bargmann space of $N$ bosons, the Laplace-Beltrami operator is defined:
\begin{equation}
\begin{split}
H_{LB}^\alpha & = \sum_{k=1}^{N} (z_k \partial_k)^2 + \frac{1}{\alpha} \sum_{i < j} \frac{z_i + z_j}{z_i - z_j}(z_i \partial_i - z_j \partial_j) \\
& = \sum_{i=1}^{N} O_i + \frac{1}{2 \alpha}  \sum_{i \neq j} V_{ij}.
\end{split}
\end{equation}
Jack polynomials can be defined, up to normalization, by the following conditions \cite{Lapointe:JC:2001}:
\begin{equation}
\begin{split}
& J_{\bm{\lambda}}^\alpha = \sum_{{\bm{\mu}} < {\bm{\lambda}}} c_{{\bm{\lambda}} {\bm{\mu}}} \textrm{m}_{{\bm{\mu}}}, \\
& H_{LB}^\alpha J_{\bm{\lambda}}^\alpha = E_{\bm{\lambda}}^\alpha J_{\bm{\lambda}}^\alpha, \\
& E_{\bm{\lambda}}^\alpha = \rho_{\bm{\lambda}}^{(B)} + \frac{1}{\alpha} (N+1)\sum \lambda_i,  
\end{split}
\end{equation}
where $\rho^{(B)}_{\bm{\lambda}}=\sum_i\lambda_i(\lambda_i - \frac{2}{\alpha} i)$ and $\textrm{m}_{\bm{\lambda}}$ are elements of the monomial basis, introduced in section \ref{sec_barg}.
Hence Jack polynomials $J_{\bm{\lambda}}^\alpha$ are the eigenvectors of the Laplace-Beltrami operator, whose decomposition in the monomial basis involves only $\textrm{m}_{\bm{\lambda}}$ and squeezed terms.\\
Notice that the basis of permanents $\textrm{m}_{\bm{\lambda}}$ was used in the definition, to recall the usual mathematical notation. The basis of $\textrm{per}_{\bm{\lambda}}$ can be used as well, with a proper rescaling of $c_{{\bm{\lambda}} {\bm{\mu}}}$ coefficients.
For example:
\begin{equation}
\begin{split}
J^{-2}_{(2,0)} (z_1, z_2) &= (z_1 - z_2)^2 = \textrm{per}_{(2,0)}(z_1, z_2) - \textrm{per}_{(1,1)}(z_1, z_2) \\
&= \textrm{m}_{(2,0)}(z_1, z_2) - 2 \textrm{m}_{(1,1)}(z_1, z_2).
\end{split}
\end{equation}
Physical interest for Jack polynomials arise both for positive and for negative $\alpha$.
The first case is related to Calogero-Sutherland models \cite{Lapointe:CMP:1996}.
The latter is related to fractional quantum Hall effect model wavefunctions. 

\subsection{Second quantization of Laplace-Beltrami operator}

The second quantization of the Laplace-Beltrami operator, in term of the bosonic operators $a$ and $a^\dagger$ is given by:
\begin{equation}
\begin{split}
H_{LB}^\alpha \! = \!\!\! \sum_{t, u = 0}^\infty \! \! \braket{t | O_1 | u} a^\dagger_t a_u \! + \! \frac{1}{2 \alpha} \!\!\!\! \sum_{s,t,r,m=0}^\infty \!\!\!\!   \braket{st | V_{12} | rm} a^\dagger_s a^\dagger_t a_m a_r.
\end{split}
\end{equation}
Notice that here $\ket{a b}$ is a factored state, not symmetrized.\\
The first matrix element evaluates to:
\begin{equation}
\begin{split}
\braket{t | O_1 | u} = \frac{1}{\sqrt{t! u!}} \int d\mu (z^*)^t (z \partial)^2 z^u = \sqrt{\frac{u!}{t!}} u^2 \delta_{t, u},
\end{split}
\end{equation}
where $d\mu = (1/\pi) d^2 z \exp[-|z|^2/2]$ is the measure in the one-particle Bargmann space.\\
The second matrix element is 0 for $u=m$. It is computed for $u>m$:
\begin{widetext}
\begin{equation}
\begin{split}
\braket{s t | V_{12} | r m} &= \frac{1}{\sqrt{s! t! r! m!}} \int d\mu_1 d\mu_2 (z_1^sz_2^t)^* \frac{z_1 + z_2}{z_1 - z_2} (z_1\partial_1 - z_2\partial_2) (z_1^r z_2^m) \\
&= \frac{r-m}{\sqrt{s! t! r! m!}} \int d\mu_1 d\mu_2 (z_1^sz_2^t)^* (z_1+ z_2) (z_1z_2)^m \sum_{k=1}^{r-m} z_1^{r-m-k}z_2^{k-1} \\
&= 2 \sqrt{\frac{s! t!}{r! m!}} (r-m) \sum_{k=1}^{r-m} \delta_{s, k+m}\delta_{t, r-k} . \\
\end{split}
\end{equation}
\end{widetext}
Thus:
\begin{equation}
\label{eq: 2qbose}
\begin{split}
H_{LB}^\alpha 
& = \sum_{r=0}^\infty r^2 \hat{n}_r 
+ \frac{1}{\alpha}  \sum_{r=1}^{\infty} \sum_{m=0}^{r-1}  (r-m) \hat{n}_m \hat{n}_r\\
& \qquad + \frac{1}{\alpha}  \sum_{r=1}^{\infty} \sum_{m=0}^{r-1}  (r-m) \sum_{k=1}^{r-m-1} s_{m, r, k}.
\end{split}
\end{equation}
The last term is a sum over all possible squeezings, identified by the destruction sites $r$ and $m$ and by the inward shift $k$. It can be simplified by setting $s=m$, $t=k$ and $u=r-m-k$, i.e. converting the sums by summing over the lowest destruction site $s$ and the relative distances of the created particles from $s$, namely $t$ and $u$. 
\begin{equation}\label{universal}
\begin{split}
H_{LB}^\alpha 
&= \sum_{r=0}^\infty r^2 \hat{n}_r 
+ \frac{1}{\alpha}  \sum_{r=1}^{\infty} \sum_{m=0}^{r-1}  (r-m) \hat{n}_m \hat{n}_r
\\
&\qquad \quad + \frac{2}{\alpha}  \sum_{s=0}^{\infty} \sum_{t=1}^{\infty} \sum_{u=1}^{\infty} u \, s_{s, s+t+u, t} \\
&= D + K S, \\ 		
\end{split}
\end{equation}
with $D$ diagonal, $S$ sum of squeezings and $K = 2/\alpha$.

\subsection{Fermionic Laplace-Beltrami operator and fermionic Jack polynomials}

Let us consider the \emph{fermionic} Jack polynomials, i.e. Jack polynomials times a Vandermonde, in the space of $N$ fermions:
\begin{equation}
\begin{split}
S_{{\bm{\lambda}}'}^\alpha(\vec{z}) &= J_{\bm{\lambda}}^\alpha(\vec{z}) \, \left[ \prod_{i<j} (z_i - z_j) \right], \\
{\lambda}'_i &= \lambda_i +N-i.
\end{split}
\end{equation}
From this definition, one can construct a \emph{fermionic} Laplace-Beltrami operator\cite{Thomale:PRB:2011}, diagonal on the fermionic Jack polynomials:
\begin{equation}
H_{LB, F}^\alpha = \sum_{i=1}^{N} O_i + \left( \frac{1}{\alpha} - 1 \right) \frac{1}{2} \sum_{i \neq j} V^F_{ij},
\end{equation}
where $O_i=(z_i \partial_i)^2$ and
\begin{equation}
V^F_{ij} = \frac{z_i + z_j}{z_i - z_j}(z_i \partial_i - z_j \partial_j) - 2\frac{z_i^2 + z_j^2}{(z_i - z_j)^2}  .
\end{equation}
It is possible to show that:
\begin{equation}
\begin{split}
& S_{{\bm{\lambda}}}^\alpha = \sum_{{\bm{\mu}} < {\bm{\lambda}}} c_{{\bm{\lambda}} {\bm{\mu}}} \textrm{sl}_{{\bm{\mu}}}, \\
& H_{LB, F}^\alpha S_{{\bm{\lambda}}}^\alpha = E_{{\bm{\lambda}}}^\alpha S_{{\bm{\lambda}}}^\alpha, \\
& E_{{\bm{\lambda}}}^\alpha = \rho_{\bm{\lambda}}^{(F)} + \frac{K}{2} \left[ (N+1) \sum_{k=1}^N \lambda_k - (N^2-N) \right]  ,
\end{split}
\end{equation}
where $\rho^{(F)}_{\bm{\lambda}} = \sum_i \lambda_i (\lambda_i + 2i (1-1/\alpha)) $ and $K = \left( \frac{2}{\alpha} - 2 \right)$.

\subsection{Second quantization of fermionic Laplace-Beltrami operator}

The second quantization of the fermionic Laplace-Beltrami operator, in term of the fermionic creation and destruction operators, is given below.
\begin{equation}
\begin{split}
H_{LB, F}^\alpha 
&=  \sum_{r} r^2 \hat{n}_r  \\
& + \frac{K}{2} \sum_{r>m} \sum_{s,t=0}^\infty \bra{st}  V_{12}^F \left( \ket{rm} - \ket{mr} \right) a^\dagger_s a^\dagger_t a_m a_r,\\
\end{split}
\end{equation}
where we used the fact that the matrix element vanishes for $r=m$ and that $a_m a_r = - a_r a_m$. \\
$V_{12}^F \left( \ket{rm} - \ket{mr} \right)$ is computed (for $r>m$):
\begin{widetext}
\begin{equation}
\begin{split}
& V_{12}^F \left( \ket{rm} - \ket{mr} \right) = \left[ \frac{z_1 + z_2}{z_1 - z_2}(z_1 \partial_1 - z_2 \partial_2) - 2\frac{z_1^2 + z_2^2}{(z_1 - z_2)^2}  \right] (z_1^r z_2^m - z_1^m z_2^r) \\ 
&= \frac{z_1^m z_2^m}{z_1-z_2} \left[ (r-m) \left( z_1^{r-m} + z_2^{r-m} \right) (z_1 + z_2) - 2 \frac{z_1^2+z_2^2}{z_1-z_2} \left( z_1^{r-m} - z_2^{r-m} \right) \right]\\
&= \frac{z_1^m z_2^m}{z_1-z_2} \left[ \sum_{l=1}^{r-m} \left[ \left( z_1^{r-m} + z_2^{r-m} \right) (z_1 + z_2) \right] -  (z_1^2+z_2^2)  \sum_{l=1}^{r-m} \left[ z_1^{r-m-l}z_2^{l-1} + z_1^{l-1}z_2^{r-m-l} \right] \right] \\
&= (z_1^m z_2^m) \sum_{l=1}^{r-m} \left[ (z_1^{r-m-l+2} - z_2^{r-m-l+2}) \sum_{t=1}^{l-1} \frac{z_1^{l-1-t} z_2^{t-1} + z_1^{t-1} z_2^{l-1-t}}{2} \right. \\
& \qquad\qquad\qquad\qquad\qquad\qquad\qquad\qquad \;\, \left. +  (z_1^{r-m-l}z_2 - z_1 z_2^{r-m-l}) \sum_{t=1}^l \frac{z_1^{l-t} z_2^{t-1} + z_1^{t-1}z_2^{l-t} }{2} \right] \\
&= (r-m-2)\textrm{sl}_{(r, m)} + 2 \sum_{l=1}^{r-m-1} (r-m-l)\textrm{sl}_{(r-l, m+l)}.
\end{split}
\end{equation}
\end{widetext}

Finally, the second quantization of the fermionic Laplace-Beltrami operator is:
\begin{equation}
\label{eq: 2qferm}
\begin{split}
H_{LB, F}^\alpha = \sum_{r}& r^2 \hat{n}_r + \frac{K}{2}  \sum_{r=1}^{\infty} \sum_{m=0}^{r-1} \left[ (r-m-2)\hat{n}_r \hat{n}_m \right] \\
&+ \frac{K}{2} \sum_{r=1}^{\infty} \sum_{m=0}^{r-1} \sum_{k=1}^{r-m-1} (r-m-2k) s_{m, r, k} , 
\end{split}
\end{equation}
In analogy with the bosonic case, we rewrite the squeezing sum as 
\begin{equation}
\begin{split}
\sum_{r=1}^{\infty} & \sum_{m=0}^{r-1} \sum_{k=1}^{r-m-1} (r-m-2k) s_{m, r, k} 
\\
& = 2 \sum_{s=0}^\infty \sum_{u=1}^\infty \sum_{u=1}^\infty u \, s_{s, s+t+u, t}. 
\end{split}
\end{equation}
Therefore:
\begin{equation}
\begin{split}
H_{LB, F}^\alpha = H_{0, F}^\alpha + V_F^\alpha = D + K S, 
\end{split}
\end{equation}
where the second equality holds apart from a constant term, $D$ and $S$ are those introduced in Eq. \eqref{universal}.

\section{Proof of Eqs.(\ref{stp2}, \ref{rec_rel_bose}, \ref{rec_rel_fermi}) }\label{AppB}

In this section the non-straightforward calculations needed to obtain Eqs.(\ref{rec_rel_bose}, \ref{rec_rel_fermi}) are presented. They are very similar to those needed to obtain Eq.\eqref{stp2}, but for two points:
\begin{itemize}
\item since $\ket{\textrm{m}_{\bm{\lambda}}}$ and $\ket{\textrm{sl}_{\bm{\lambda}}}$ are unnormalized states, in the proof of Eqs.(\ref{rec_rel_bose}, \ref{rec_rel_fermi}) there is a coefficient $\nu_{\bm{\mu}} = \braket{{\bm{\mu}} | {\bm{\mu}}}$;
\item when acting with the squeezing operator on permanents or monomials, we obtain different coefficients, as pointed out in Eqs.\eqref{sonper} and \eqref{sonmon}. This lead to the difference between Eqs.\eqref{ydefper} and \eqref{ydefmon}. 
\end{itemize}
In the following, only the monomial case is considered. 
 
We now prove the following equality, where $\ket{{\bm{\lambda}}} = \ket{\textrm{m}_{\bm{\lambda}}}$ for bosons, $\ket{{\bm{\lambda}}} = \ket{\textrm{sl}_{\bm{\lambda}}}$ for fermions:
\begin{equation}
\frac{1}{\nu_{\bm{\mu}}} \! \sum_{{\bm{\mu}}<{\bm{\mu}}'<{\bm{\lambda}} } \!\!\! \bra{{\bm{\mu}}} S \ket{{\bm{\mu}}'}  \mathsf{b}_{{\bm{\lambda}},{\bm{\mu}}'} \!=\! \!\! \sum_{{\bm{\theta}}; {\bm{\mu}} < {\bm{\theta}}}\! \genfrac{}{}{0pt}{1}{(\theta_i-\theta_j)}{(\mu_i - \mu_j)} \mathsf{y}_{\mu_i,\mu_j} (\pm 1)^{N_\mathrm{sw}} \mathsf{b}_{{\bm{\lambda}}{\bm{\theta}}},
\end{equation}
where $\mathsf{y}_{\mu_i, \mu_j}$ is defined in Eq.\eqref{ydefmon}.
Using $\nu_{\bm{\mu}} = 1$ and normalized $\ket{{\bm{\lambda}}}$, this same proof accounts for equation \eqref{stp2}.
First, the operator $S$ must be rewritten to recast the sum over all the possible squeezings into a sum over squeezed partitions.
It was shown in Eq.(\ref{eq: 2qbose}) and in Eq.(\ref{eq: 2qferm}) that
\begin{equation}
\begin{split}
S^{(B)} &= \frac{1}{2}  \sum_{r=1}^{\infty} \sum_{m=0}^{r-1}  (r-m) \sum_{k=1}^{r-m-1} s_{m, r, k} \qquad \; \, \,  \text{for bosons,} \\
S^{(F)} &=  \frac{1}{2} \sum_{r=1}^{\infty} \sum_{m=0}^{r-1} \sum_{k=1}^{r-m-1} (r-m-2k) s_{m, r, k} \ \, \text{for fermions.} \\
\end{split}
\end{equation}
For bosons, we have:
\begin{equation}
\begin{split}
& S^{(B)} = \frac{1}{2}  \sum_{r=1}^{\infty} \sum_{m=0}^{r-1}  (r-m)  \left[ \sum_{k=1}^{\lfloor (r-m)/2 \rfloor} s_{m, r, k} \right. \\
& \qquad \qquad \qquad \left. +\sum_{k=\lceil (r-m)/2\rceil}^{r-m-1} s_{m, r, k} + s_{m, r, (r-m)/2}\right] \\
& = \frac{1}{2}  \sum_{r=1}^{\infty} \sum_{m=0}^{r-1}  (r-m)  \left[ 2 \sum_{k=1}^{\lfloor (r-m)/2 \rfloor} s_{m, r, k} + s_{m, r, (r-m)/2} \right] ,\\
\end{split}
\end{equation}
where the terms with $s_{m, r, (r-m)/2}$ are present only if \mbox{$(r-m)/2 \in \mathbb{N}$}, $\lfloor x \rfloor$ is the greatest integer number smaller than $x$, $\lceil x \rceil$ is the smallest integer number greater than $x$. Here the fact that $s_{u, m, k} = s_{u, m, m-u-k}$ has been used.
Then:
\begin{equation}
\begin{split}
& S^{(B)} \ket{{\bm{\lambda}}} \\
&= \frac{1}{2}  \sum_{r=1}^{\infty} \sum_{m=0}^{r-1}  (r-m)  \left[ 2 \sum_{k=1}^{\lfloor (r-m)/2 \rfloor} (n_{m+k}+1) (n_{r-k}+1) \right. \\
& \left. \qquad\qquad\qquad\qquad\qquad\qquad\qquad\quad + 2 \binom{n_{(r+m)/2}}{2}  \right] \ket{{\bm{\mu}}}  \\
&= \sum_{{\bm{\mu}} \leftarrow {\bm{\lambda}}} (\lambda_r - \lambda_m) \mathsf{y}_{\mu_r, \mu_m} \ket{{\bm{\mu}}}, 
\end{split}
\end{equation}
where the sums over positions involved in the squeezing are converted in a sum over squeezed partitions ${\bm{\mu}}$, $\lambda_r$ and $\lambda_m$ are the quantum numbers of the annihilated particles and $\mu_r$ and $\mu_m$ are those of the created particles. Notice that ${\bm{\mu}}$ depends on $r, m, k$.

For fermions, we have:
\begin{equation}
\begin{split}
S^{(F)} 
&= \frac{1}{2}  \sum_{r=1}^{\infty} \sum_{m=0}^{r-1}    \left[ \sum_{k=1}^{ \lfloor (r-m)/2 \rfloor} (r-m-2k) s_{m, r, k} \right. \\
&\qquad \qquad \qquad \left. + \sum_{k=\lfloor (r-m)/2\rfloor}^{r-m-1} (r-m-2k) s_{m, r, k} \right] \\
&= \sum_{r=1}^{\infty} \sum_{m=0}^{r-1}   \left[  \sum_{k=1}^{ \lfloor (r-m)/2 \rfloor} ((r-k) - (m+k)) s_{m, r, k}\right] ,\\
\end{split}
\end{equation}

since in the fermionic case the creation of two particles in site $(r+m)/2$ is forbidden due to the Pauli principle. Here the fact that $s_{u, m, k} = - s_{u, m, m-u-k}$ has been used.
Then:
\begin{equation}
\begin{split}
S^{(F)} \! \ket{{\bm{\lambda}}} \! & = \! \! \sum_{r=1}^{\infty} \! \sum_{m=0}^{r-1} \! \left[ \!  \sum_{k=1}^{ \lfloor (r-m)/2 \rfloor} \!\! ((r-k) \! - \! (m+k)) (-)^{N_\mathrm{sw}} \! \ket{{\bm{\mu}}} \! \right]\\
&= \sum_{{\bm{\mu}} \leftarrow {\bm{\lambda}}} (\mu_r - \mu_m) (-)^{N_\mathrm{sw}} \ket{{\bm{\mu}}},
\end{split}
\end{equation}
in complete analogy with the bosonic case. Notice that the $\mathsf{y}$ factor for fermions always equals 1.

Finally:		
\begin{equation}
\begin{split}
\frac{1}{\nu_{\bm{\mu}}}  & \sum_{{\bm{\mu}} < {\bm{\theta}} < {\bm{\lambda}}} \braket{{\bm{\mu}}|S^{(B/F)}|{\bm{\theta}}} \mathsf{b}_{{\bm{\lambda}}, {\bm{\theta}}} \\
& \!\!\!\! = \frac{1}{\nu_{\bm{\mu}}}  \sum_{{\bm{\mu}} < {\bm{\theta}} < {\bm{\lambda}}} \! \bra{{\bm{\mu}}} \! \sum_{{\bm{\mu}}' \leftarrow {\bm{\theta}}}  \genfrac{}{}{0pt}{1}{(\theta_i-\theta_j)}{({\bm{\mu}}'_i - {\bm{\mu}}'_j)} \mathsf{y}_{\mu'_i, \mu'_j} (\pm 1)^{N_\mathrm{sw}} \ket{{\bm{\mu}}'} \mathsf{b}_{{\bm{\lambda}}, {\bm{\theta}}} \\
& \!\!\!\! = \sum_{{\bm{\mu}} \leftarrow {\bm{\theta}} < {\bm{\lambda}}} \genfrac{}{}{0pt}{1}{(\theta_i-\theta_j)}{(\mu_i - \mu_j)} \mathsf{y}_{\mu_i, \mu_j} (\pm 1)^{N_\mathrm{sw}} \mathsf{b}_{{\bm{\lambda}}, {\bm{\theta}}}.
\end{split}
\end{equation}

\vspace{3em}
$ $\\
\vspace{3em}

\bibliographystyle{apsrev4-1}

\bibliography{BibliographyFQ}

\end{document}